\documentclass[%
reprint,
nofootinbib,
 amsmath,amssymb,
aps,
pre,
]{revtex4-2}

\usepackage{graphicx}
\usepackage{dcolumn}
\usepackage{bm}

\usepackage{booktabs}
\usepackage{multirow}
\usepackage{tikz}
\usepackage{pgfplots}
\pgfplotsset{compat=1.18}
\usetikzlibrary{fadings,arrows.meta}
\usetikzlibrary{shapes.misc}
\tikzset{cross/.style={cross out, draw=black, minimum size=2*(#1-\pgflinewidth), inner sep=0pt, outer sep=0pt},
cross/.default={1pt}}
\tikzfading[name=fade out, inner color=transparent!0, outer color=transparent!100]
\tikzfading[name=fade right, left color=transparent!0, right color=transparent!100]

\definecolor{mycolor1}{rgb}{0.55,0.75,0.875}%

\usetikzlibrary{external}
\tikzset{external/system call={lualatex \tikzexternalcheckshellescape -halt-on-error -interaction=batchmode -jobname "\image" "\texsource"}}
\tikzset{external/only named=true}

\usepackage[ruled,vlined]{algorithm2e}

\usepackage{siunitx}

\newcommand{\R}{\ensuremath\mathbb{R}}

\newcommand{\eg}{{\it e.g.}}

\makeatletter
\let\div\@undefined                        
\makeatother
\DeclareMathOperator{\div}{div}

\begin{document}

\preprint{PRE}

\title{A PDE Model for Janus Particle Swarming}
\thanks{This work was partially supported by NIGMS of the NIH (under Award Number P20GM103474), the Undergraduate Scholars Program (USP) and the Department of Mathematical Sciences at Montana State University, and a collaboration grant for mathematicians by the Simons Foundation (Award Number 586942). S.G.M. would like to acknowledge the support of NSF grants $\#1813654$, $\#2112085$ and the Army Research Office (W911NF-19-1-0288).}

\author{Griffin Smith}

\author{Nathan Stouffer}

\author{Scott G. McCalla}
\email{scott.mccalla@montana.edu}
\affiliation{Department of Mathematical Sciences, Montana State University}

\author{Dominique P. Zosso}
\email{Dominique.zosso@zootweb.com}
\affiliation{Zoot Enterprises, Inc.}

\date{\today}

\begin{abstract}
    It has been experimentally shown that Brownian motion and active forward drift controlled by quorum sensing is sufficient to produce clustering behavior in orientable Janus particles. This paper explores the group formation and cohesion of Janus particles using both an agent-based computer model and a nonlocal advection-diffusion partial differential equation (PDE) model. Our PDE model can recreate the behavior from both physical experimentation and computer simulations. Additionally, the PDE model highlights an annular structure which was not apparent in prior work.
\end{abstract}


\maketitle

\section{Introduction}
\label{sec:introduction}

Systems of individual particles governed by simple nonlocal interactions often exhibit complex emergent behavior: these swarms appear to act in concert as if controlled by a single entity. Flocks of birds~\cite{vicsek1995novel,cucker2007emergent,van2015collective}, schools of fish~\cite{huth1992simulation,shinchi2002fractal}, swarms of insects~\cite{topaz2008model,SwarmPrimer,Topaz:2012kg,reid2015army}, zebrafish stripe patterns~\cite{volkening2015modelling,volkening2018iridophores,volkening2020modeling} predator-prey systems~\cite{chen2014minimal} and even crowds of people~\cite{helbing2001self,bertozzi2015contagion,wang2017efficient} are all examples of such emergent behavior. While many instances of the relationship between simple interactions and the induced coherent structures have been studied in detail from a theoretical standpoint, there is a dearth of experimental systems where the underlying interactions are known and where the collective behavior of the model can be compared to the experimental results.  

Typically, these models depend on pairwise forces between particles in the system.  These forces might depend on the distance between particles~\cite{motsch2011new}, the relative orientation of nearby particles~\cite{vicsek1995novel}, or a combination of these variables~\cite{huth1992simulation,shinchi2002fractal,evers2015anisotropic,von2016anisotropic,o2017oscillators,o2018ring}.  Directly analyzing these discrete particles systems is often intractable, but detailed information about the discrete coherent dynamics can be gained by studying partial differential equations (PDEs) for the particle density function $\rho(\theta,\Vec{x},t)$ that arise in the large particle number limit~\cite{mogilner1998nonlocal,Kolokolnikov:2010,eftimie2012hyperbolic,vonBrecht:2012a,vonBrecht:2012b,LevyCrime,mccalla2018consistent,degond2008continuum,bertozzi2015ring}.  These PDE models often provide predictive information for these systems even when the number of particles is relatively small.

In \cite{lavergne2019group}, an experimental system of active Janus particles---spherical particles that are divided into a light and dark hemisphere at the equator---demonstrated that limited perception and activation can lead to the formation of a cohesive group. The system consists of $n$ randomly moving particles in a flat plane, each with an orientation determined by the alignment of their light and dark halves. In the experiment real-time image analysis was used to find each particle's position and orientation, and then a vision cone was computed for each particle; see fig.~\ref{fig:model}.  A weighted sum of the number of perceived particles in each vision code is then calculated.  When this sum exceeds a given threshold for a particular particle, a laser is used to activate this particle driving it in the forward direction.  Surprisingly, these Janus particle experiments established that combining this diffusion and nonlocal activation was enough to cluster and form patterned states.  This is in contrast to models that induce group structure by active alignment and attraction.  This experimental study was complemented with an agent-based model which reproduced this clustering behavior.  The authors were then able to identify parameter regimes that induced group formation and estimate various bifurcation parameters with respect to the angle of the vision cone and the activation threshold of  particles.

In the current study, we seek to complement \cite{lavergne2019group} with extensive numerical simulations on an agent-based model and a related large particle number limit advection-diffusion PDE. The PDE models the evolution of the particle density function, $\rho(\theta,\Vec{x},t)$, which depends on alignment, position in space, and time.  Our numerical simulations are computationally cheap and allow for a detailed study of the clustered states over a wide range of parameter values allowing us to understand the attracting states in the experimental system.  The validity of the PDE model is confirmed against the agent based model and allows for a detailed picture of the attracting profiles' shape, height, and width without the noise inherently present in both the experimental system and the agent based model.  The simulations show $\rho$ evolving towards a localized annular structure for large regions in parameter space; the density looks like it collapses on a round disc with a noticeable peak in the density at the periphery of the disc.  This behavior is confirmed with the agent based model.
This annular structure to the groups further elucidates the coherent behavior seen in \cite{lavergne2019group} and was not noted in the original publication. This localized pattern's parameter dependence and bifurcation structure is then extensively explored using the PDE model.

\section{Methods}
\label{sec:methods}

\begin{figure}
\centering\includegraphics[page=1]{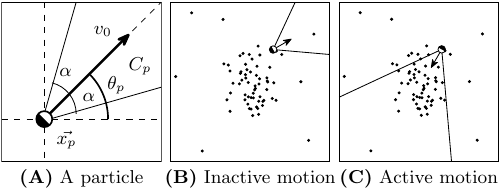}
    \caption{\textbf{(A)} A Janus particle with orientation $\theta_p$, associated vision cone $C_p$ defined by the half-angle $\alpha$, and activated forward velocity $v_0$. \textbf{(B)} An outward facing, inactive Janus particle on the edge of a cluster; inactive particles are moving subject to Brownian motion (position and orientation). \textbf{(C)} An inward facing, active Janus particle on the edge of a cluster; in addition to Brownian motion, active particles move forward with velocity $v_0$.}
    \label{fig:model}
\end{figure}

The experimental system explored in \cite{lavergne2019group} consists of $n$ particles at positions $\vec{x}_p \in \mathbb{R} ^2$ for $p=1,\ldots,n$ each with orientation $\theta_p$.
Each agent has an associated vision cone $C_p$ of half-angle $\alpha$ measured from $\theta_p$ with its tip at $\vec{x}_p$: particles can sense other particles within their vision cone. The perception weight is defined as 
\begin{equation} \label{eq:weights}
w (p, q) = 
\begin{cases}
    1/d(\vec{x}_p,\vec{x}_q) & \text{if } \vec{x}_q \in C_p \setminus \{\vec{x}_p\}\\
    0 & \text{otherwise,}
\end{cases}
\end{equation}
with $d(\vec{x}_p,\vec{x}_q)$ the Euclidean distance between agents $p$ and $q$.  For a particle $p$, its perception is defined as $\sum_{q} w(p,q)$. Note that this is an asymmetric model of perception.
Each particle is subject to Brownian motion ($D_{\vec{x}}=\SI{0.02}{\micro\meter\squared\per\second}$), rotational diffusion ($D_{\theta}=\SI[parse-numbers = false]{1/110}{\radian\squared\per\second}$), and nonlinear, nonlocal advection. When a particle's perception exceeds an activation threshold  $P^\ast$, then $p$ is active and moves forward with speed $v_0=\SI{0.2}{\micro\meter\per\second}$. This experimental model is illustrated in Figure~\ref{fig:model}.
Here, we study this system with two approaches:
an agent based approach from~\cite{lavergne2019group} (\S\ref{sec:agentbased}) and an advective-diffusion PDE approach (\S\ref{sec:pdebased}). The code for both models can be found at \protect\url{https://github.com/nathanstouffer/janus-particles}.

\subsection{Agent-based simulation}\label{sec:agentbased}

The agent-based simulation was implemented in Java and is run through python scripts.
The simulator allows for control over all relevant parameters. Positions, orientations, and the activation status of each particle can be recorded in log-files at any time-point in the simulation. Each discrete time step of the agent-based simulation corresponds to $\Delta t= \SI{0.5}{s}$, to match the experimental setup of \cite{lavergne2019group}. Initially, agents are randomly distributed throughout the \SI{250}{\micro\meter} square and the simulation runs for a pre-determined number of steps. In the absence of particular run-time optimization, the time-complexity of each iteration is $\mathcal{O}(n^2)$, since every agent must process the other $n-1$ agents for perception-activation evaluation. The pseudo-code of the simulator is given in algorithm~\ref{alg:agentbased}. Here, $\hat{\theta}=(\cos{\theta},\sin{\theta})^\top$, $\eta_{\vec{x}}$ denotes a sample drawn from a 2D zero-mean unit-variance normal distribution, and $\eta_{\theta}$ denotes a sample drawn from a 1D zero-mean unit-variance normal distribution.

\begin{algorithm}
\SetAlgoLined
\KwResult{Position $\vec{x}^t$, orientation $\theta_p^t$, and activation $S_p^t$ of particles through time}
  \For{$p=1:n$}{
     randomly initialize $\vec{x}_p^1\in[0,1]^2$, $\theta_p^1\in[0,2\pi)$, uniformly\;
     }
  \For{$t=1:T$}{
   \For{$p=1:n$}{
      \eIf{$\sum_{q\neq p} w(p,q)>P^\ast$}{
       $S_p^t = 1$\;
      }{
       $S_p^t = 0$\;
      }
       $\vec{x}_p^{t+1} = \vec{x}_p^t + \sqrt{2D_{\vec{x}}\Delta t}\eta_{\vec{x}} + S_p^tv_0\hat{\theta}_p^t\Delta t$\;
       $\theta_p^{t+1} = \theta_p^t + \sqrt{2D_{\theta}\Delta t}\eta_{\theta} \pmod{2\pi}$\;
   }
   }
\caption{Agent-based simulation}
\label{alg:agentbased}
\end{algorithm}

The agent-based simulation uses a relatively small number of agents to match the experiments in \cite{lavergne2019group}. This results in rough estimates of spatial densities at a given time point. To improve this estimate of the spatial density and long-term behavior, we partition space into square bins and histogram the number of particles in each bin.  These histograms are then averaged over time.
This significantly improves the density estimation and  allows for comparison with the PDE model.

\subsection{Advection-Diffusion PDE}\label{sec:pdebased}
To complement the agent-based simulation, we use a nonlinear, nonlocal advective PDE that approximates the limit $n\to\infty$ for a large number of particles. We consider a density function that depends on particle orientation, time, and space:
\begin{equation}
\begin{aligned}
    \rho\colon [0,2\pi)\times \mathbb{R}_{\geq 0} \times \mathbb{R}^2 &\to \mathbb{R}\\
    (\theta,t,\vec{x}) &\mapsto \rho(\theta,t,\vec{x}).
\end{aligned}
\end{equation} 
At any point in time and space, the total density of particles, $\bar{\rho}$, across all orientations, is obtained by integration in the angular direction:
\begin{equation}
    \bar{\rho}(t,\vec{x}) := \int_0^{2\pi} \rho(\theta,t,\vec{x}) d\theta.
\end{equation}

The PDE then captures two main terms: spatial/angular diffusion and perception-based active motion.
The spatial distribution evolves according to an advection-diffusion equation, similar to \cite{mogilner1998nonlocal}, where the diffusion accounts for the particles' Brownian motion in space and orientation. Let $D_{\vec{x}}$ and $D_{\theta}$ denote the spatial and angular diffusion coefficients, respectively. 
The diffusive model then takes the following form:
\begin{equation}\rho_t = D_{\vec{x}} \Delta \rho + D_{\theta} \rho_{\theta \theta}.\end{equation}

To determine whether a particle at position $\vec{x}$ and oriented in direction $\theta$ is activated, we first define a vision kernel, $K_\theta$, based on perceptive weights in \eqref{eq:weights}, using a cone $C_0$ at the origin, with axis pointing in direction $\theta$ and half-angle $\alpha$:
\begin{equation}
    K_{\theta}(\vec{s}) := w(0,-\vec{s}).
\end{equation}
(This resulting kernel will be a perceptive cone ``looking'' in the direction opposite to $\hat{\theta}$).
We then determine perception $V(\theta,\vec{x},t)$ by performing convolution between $K_{\theta}$ and the total local particle density, $\bar{\rho}$:
\begin{equation}\label{eq:convolution}
   V(\theta,\vec{x},t) := K_{\theta}*\bar{\rho} = \iint_{\R ^2} K_\theta(\vec{x} - \vec{y}) \bar{\rho}(\vec{y},t) d\vec{y}.
\end{equation}
Activation is then represented as a binary function $S$:
\begin{equation}\label{eq:thresholding}
    S( \theta,\vec{x},t)\equiv \begin{cases}1&\text{if } V(\theta,\vec{x},t) > P^\ast,\\ 0&\text{otherwise}.\end{cases}
\end{equation}
The purely advective PDE then takes the form
\begin{equation}
   \rho_t = -\vec{\nabla}\cdot\left( Sv_0\hat{\theta}\rho\right) 
\end{equation}
and the complete advection-diffusion PDE of interest becomes
\begin{equation}
    \label{eq:pde}
    \rho_t = D_{\vec{x}} \Delta \rho  - \vec{\nabla}\cdot\left( Sv_0\hat{\theta}\rho\right)  + D_{\theta} \rho_{\theta \theta}.
\end{equation}

\subsection{Numerical solution to PDE}
For the numerical solution, we restrict the spatial domain to a square of side length \SI{250}{\micro\meter} (the experimental work in \cite{lavergne2019group} used a circular area of interest of radius $R_0=\SI{106}{\micro\meter}$). The introduction of domain boundaries requires the definition of appropriate boundary conditions. In the agent-based model, reflecting boundary conditions are implemented using simple coordinate truncation. In the PDE model, we achieve the same effect by using appropriate mirror boundary conditions (no flow across the boundary). Naturally, periodic boundary conditions apply in the angular direction.

Discretization in space of \eqref{eq:pde} brings the problem into the form of a non-linear first order ODE, $\frac{dy}{dt} = F(t,y)$, for which we can use common Runge--Kutta-type solvers with adaptive time-stepping. (The \texttt{ode45}-solver in MATLAB proved efficient for our purposes.)
For discretization, we divide the domain into $N_{\vec{x}}\times N_{\vec{x}}$  square bins (\eg, $256\times256$), spatially, and consider $N_\theta$ discrete orientations (\eg, $N_\theta=30$). We create both forward ($D_f$) and backward ($D_b$) spatial gradient operators with Neumann boundary conditions. 
The corresponding divergence operators are obtained as matching negative adjoints and the spatial Laplacian is a centered difference stencil. 
Similarly, we use a standard 3-point stencil for the periodic Laplacian in the angular direction. Activation is computed using convolution \eqref{eq:convolution} and thresholding \eqref{eq:thresholding}. Since the vision kernel is large, exploiting the convolution theorem is advantageous. We precompute the 2D FFT of all kernels, then at each update step we compute the 2D FFT of $\bar{\rho}$ with appropriate zero-padding. The activation is then obtained using point-wise spectral multiplication and inverse 2D FFT. Finally, for stability of the advective term, we use an upwind scheme: for each orientation $\theta$, we choose forward/backward divergence operators based on the quadrant containing $\hat{\theta}$.
Overall, despite employing the convolution theorem, computing the perception based on \eqref{eq:convolution} with the vision kernel $K_\theta$ at each time step and for each orientation is the most time-consuming computational step.

\section{Results}
\label{sec:results}
\begin{figure*}
\centering\includegraphics[page=2]{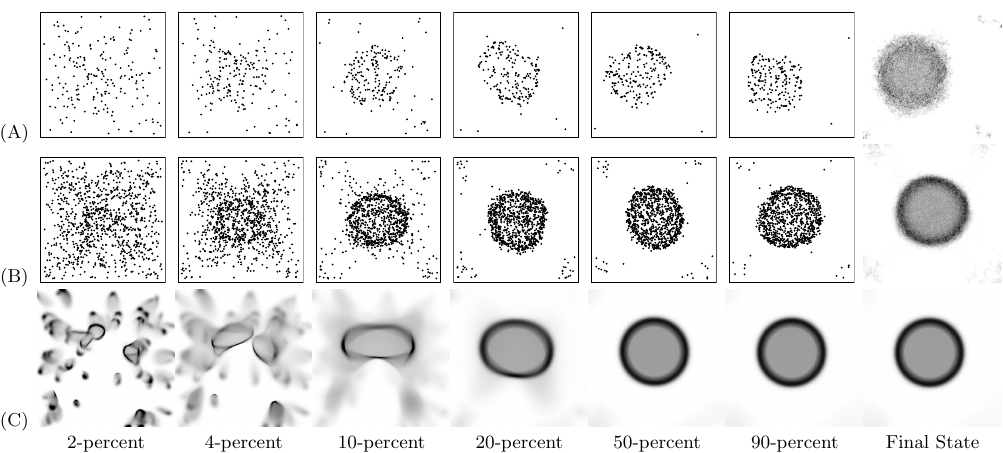}
    \caption{Simulations for nominal parameter choices $\alpha=\frac{\pi}{2}$ and $P^\ast=P_\alpha^c$. \textbf{(A)}--\textbf{(B)} Agent-based simulation with 200 and 1000 particles, respectively. 
    The final image in each row shows the aggregated histogram from  \SIrange{5000}{10000}{\second}. 
    \textbf{(C)} PDE-based simulation ($\bar\rho$) with spatial resolution: $512\times512$, angular resolution: 60, simulation time: \SI{10000}{\second}.}
    \label{fig:time-slices.1}
\end{figure*}

We ran PDE simulations at a spatial resolution of up to $512 \times 512$ points and an angular resolution of up to $60$ orientations, for up to \SI{100000}{\second} (simulated, not CPU time). On standard computing equipment and for these settings, simulation is typically faster than real time. Also, within standard floating point precision, the simulations exhibit conservation of mass. For comparison, we ran agent-based model simulations with up to \num{5000} particles. To more directly compare results from the PDE model against the agent-based model, we create histograms of the agents by aggregating particle counts in bins over time, rather than just considering the end-point. Our results from the PDE simulations show a clear ring-like structure which was not mentioned in \cite{lavergne2019group}; this annular structure is not immediately apparent in agent-based simulation end-point results. By aggregating the agent-densities over time, however, the annular structure becomes very apparent in the agent-based model as well. 

Our approach to understanding this system is as follows: we start with detailed simulations with the default parameters originating from \cite{lavergne2019group}, then investigate the apparent annular structure and active/inactive particle distributions. We then focus on demonstrating convergence to this annular pattern and consistency between our PDE and agent-based models, before exploring a wider range of model parameters using the PDE. The main parameters to explore are the vision cone angle, $\alpha$, and the perception threshold, $P^\ast$. For the latter, it is helpful to use the reference threshold $P_\alpha^c=\frac{\alpha N}{\pi^2R_0}$, as defined in \cite{lavergne2019group}.

\subsection{Annular structure for $\alpha = \frac{\pi}{2}$, $P^\ast=P^c_\alpha$}
In \cite{lavergne2019group}, disk-shaped group formation and cohesion was observed for moderate perception parameter settings $\alpha = \frac{\pi}{2}$, $P^\ast=P^c_\alpha$.
The results of two agent-based simulations (with 200 and 1000 particles, respectively) and a PDE simulation, both with the above perception parameters are shown in figure~\ref{fig:time-slices.1}. 

 \begin{figure}[t]
\centering\includegraphics[page=3]{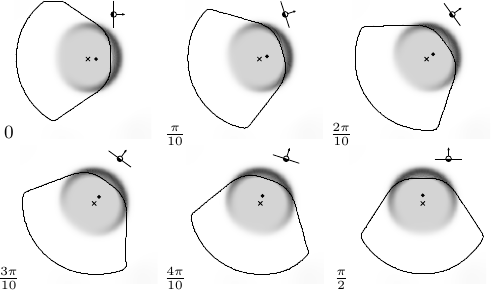}
    \caption{Particle densities $\rho(\theta)$ of selected orientations $\theta$ from the end-state of a PDE simulation. Particle orientation is indicated numerically by the angle $\theta$ (bottom left) and visually (top right). The `x' denotes the center of mass of the entire particle-density, while the `$\cdot$' denotes the orientation-specific center of mass. The solid line outlines the region within which particles of the selected orientation are active (visual perception above threshold). }
    \label{fig:angular}
\end{figure}

Starting from uniform random placement and orientation of the particles in both the agent-based and PDE models, group formation typically occurs within the first \SI{2000}{\second} of the simulation (corresponding to the first \SI{20}{\percent} of the total simulated time). These groups are cohesive: for the remaining \SI{80}{\percent} simulation time, particle densities are essentially stationary,  which suggests that observed final-states are indeed pseudo-steady states of the system. 

\subsubsection{Annular structure}
In addition to the simple disk-shaped appearance of the particle group in the agent-based models, the agent-based histogram and particularly the PDE model's density function exhibit a pronounced accumulation of particles on the periphery of the disk, resulting in the overall more annular density pattern. This is apparent in the agent-based histograms and PDE simulations in figure~\ref{fig:time-slices.1}.

\subsubsection{Orientation}
In figure \ref{fig:angular}, a final state of a PDE simulation was broken up over specific particle orientations. We note that each group of particles with the same orientation has medium density over most of the disk and an arc of higher density at the periphery of the disk where particles are ``looking'' away from the cluster. As a result, the density of particles is shifted away from the overall center of mass in the direction that particles are looking. 
Observe also that the arc of highest density is mostly excluded from the area where particles are activated. This leads to the conclusion that, per orientation and as a whole, agents along the boundary are mostly outward looking and inactive, while the agents in the interior are mostly active and travelling to the periphery.

 \begin{figure}[b]
\centering\includegraphics[page=4]{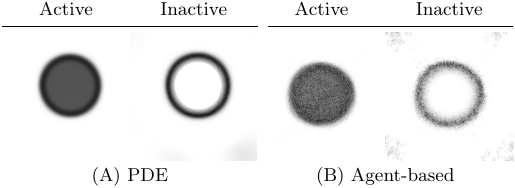}
     \caption{Comparison of active versus inactive particle densities: \textbf{(A)} from the end-state of PDE simulation, and \textbf{(B)} from histogram of agent-based simulations.} 
     \label{fig:Activation-Comparison}
 \end{figure}

\subsubsection{Activation}
Figure \ref{fig:Activation-Comparison} makes the relative distribution between active and inactive particles more obvious:
For the PDE model, the orientation-specific density was partitioned into active versus inactive using the activation boundary seen in figure~\ref{fig:angular}, then aggregated over all orientations. This results in the two density maps for active versus inactive particle densities, seen in figure~\ref{fig:Activation-Comparison}. Similarly, it is straightforward to compute aggregate densities for active and inactive particles directly from the agent-based model simulations. The two models agree extremely well in the observations: inactive particles almost exclusively occur in the annular region, while a very small fraction of particles is stuck in the corners of the simulation domain. The interior of the disk contains only active particles. 

\subsubsection{Symmetry} 
\begin{figure*}
\centering\includegraphics[page=5]{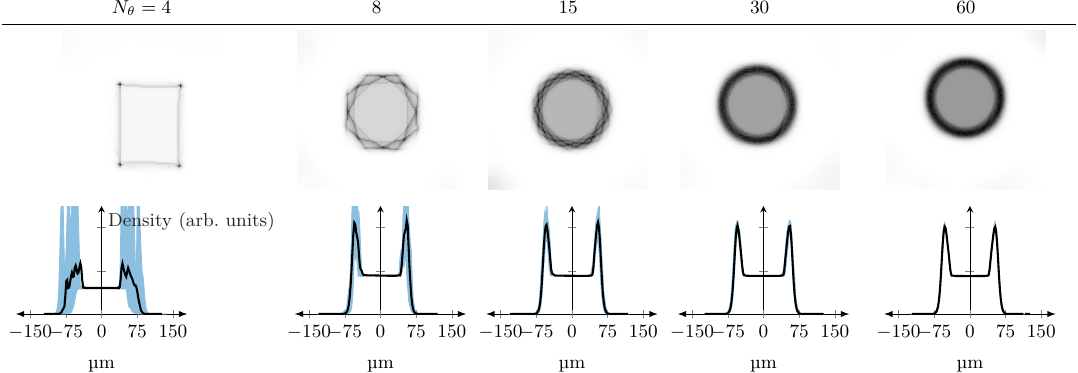}
\caption{PDE simulations for varying angle discretizations $N_\theta$. \textbf{Top}: End-state $\bar{\rho}$ of the PDE simulation. \textbf{Bottom}: Centered cross-sections at 20 different angles. Average cross-section (solid black) and \SIrange{5}{95}{\percent} confidence interval (blue shade). (18 out of 20 profiles are contained within the blue shaded region.) }
\label{fig:AngConvergence}
\end{figure*}

Next, we examine the symmetry of the annular structure. In figure~\ref{fig:AngConvergence}, we observe how the model behaves at different angular PDE discretizations and compare density profiles as cross-sections of different angles through the disk's center of mass.
An annular pattern appears for all chosen angular resolutions $N_\theta$ (the number of different orientations for particles in the simulation). Not surprisingly, an insufficient angular resolution greatly impacts the shape of the observed annulus. In the extreme case of just four possible particle orientations, the end-state resembles a set of four clusters at the corners of a square, with an exchange of particles between them mostly happening on jets along the edges of the square. Medium resolutions better approximate a circular shape, albeit with strong resolution artifacts present. Above an angular resolution of \numrange{30}{60} different orientations, however, the annulus exhibits near perfect rotational symmetry. As the number of angles increases the confidence interval of profiles (shaded region) collapses to the mean cross-section until the two are virtually indistinguishable. While at a high enough angular discretization, any angled cross-section through the center of mass is representative of the averaged cross-section. As a result we will base further cross-sectional analysis on the mean profile for denoising purposes.

\subsection{Model consistency and convergence}
We now use the averaged density cross-sections for a visual comparison of consistency between the agent-based and PDE models, as well as for convergence of the PDE model with respect to time and spatial resolution. The results are illustrated in figure~\ref{fig:cross-sections}.

\subsubsection{Model consistency}
First, we compare the agent-based model densities (using time-aggregated histograms) against a PDE end-state simulated using the same model parameters. 
In figure~\ref{fig:cross-sections}~\textbf{(A)}, the thickened blue curve represents the PDE cross-section of the final state and is compared against the cross-sections from the histograms of various agent-based simulations. The agent-based simulations were run at the same model parameters as the PDE simulation only with varying particle numbers. The histograms were aggregated at a spatial resolution of $512\times512$ to match the spatial resolution of the PDE simulation. For comparison the histograms were normalized to match the mass of the PDE simulation, and centered cross-sections angularly averaged for noise reduction---the agent-based cross-sections still appear less smooth due to the stochastic nature of the agent-based model. As the agent-based simulation's particle number increases, the relative size of the stochastic noise decreases, and the cross-sections gradually approach the PDE cross-section, indicating that the PDE model does indeed represent a continuum limit of the agent-based model.

\subsubsection{Temporal convergence}
Next, we look at temporal convergence of the PDE model towards a supposed steady state. Again, taking cross-sections through the annulus' center of mass shows the structure quite clearly. In figure~\ref{fig:cross-sections}~\textbf{(B)}, we plot averaged cross-sections of the PDE at 18 different points in time. The progression through time clearly indicates a convergence from a discrete random distribution to an annular structure.

\subsubsection{PDE discretization}
\begin{figure*}
\centering\includegraphics[page=6]{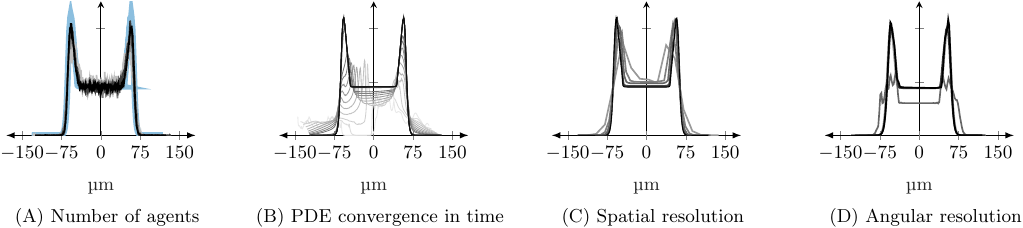}
\caption{Profile cross-sections for consistency and convergence analysis. \textbf{(A)} Agent-based model vs.\ PDE. Plotted increasingly darker in gray-scale are the centered and averaged cross-sections of four agent-based simulation histograms with particle counts of 500, 1000, 2000, and 5000, respectively. Cross-sections of the final state of the PDE model are shown in thick blue. \textbf{(B)} PDE convergence in time. 
Here, as time increases, the cross-section takes on a darker hue. \textbf{(C)} Plotted increasingly darker in gray-scale are the averaged cross-sections of the end states of the PDE model run at a spatial resolution $N_{\vec{x}}$ of 16, 32, 64, 128, 256, and 512, respectively. \textbf{(D)} Increasing angular resolution $N_\theta$: the averaged cross-sections of figure~\ref{fig:AngConvergence} are overlaid with darker hues as the resolution improves. }
\label{fig:cross-sections}
\end{figure*}

In figure~\ref{fig:cross-sections}~\textbf{(C)}, we examine results from PDE simulations at different spatial resolutions $N_{\vec{x}}\times N_{\vec{x}}$. 
We note that simulations are consistent over large stretches of resolution choices: increased spatial resolution allows for sharper peaks and overall cleaner distributions, while the lowest resolution simulations exhibit substantial discretization noise.

Finally, in figure~\ref{fig:cross-sections}~\textbf{(D)} we illustrate in the same manner the convergence of the averaged cross-sections as the angular discretization $N_\theta$ is changed (results from figure~\ref{fig:AngConvergence}). Predictably, using 4 angles yields poor results.  As the angular resolution is increased, the annular limit quickly emerges.

Overall, our results indicate strong consistency between the agent-based and the PDE model at reasonable spatial and angular discretization, such as $N_{\vec{x}}=256$ and $N_\theta=30$.

\subsection{Pattern formation for varying parameters}
We now turn to examining the parameter space defined by the perception angle $\alpha$ and the activation threshold $P^\ast$.
For comparison to results in \cite{lavergne2019group}, we control the threshold $P^\ast$ in terms of the reference threshold $P^c_\alpha$ defined at the beginning of this section.

\begin{figure*}
\centering\includegraphics[page=7]{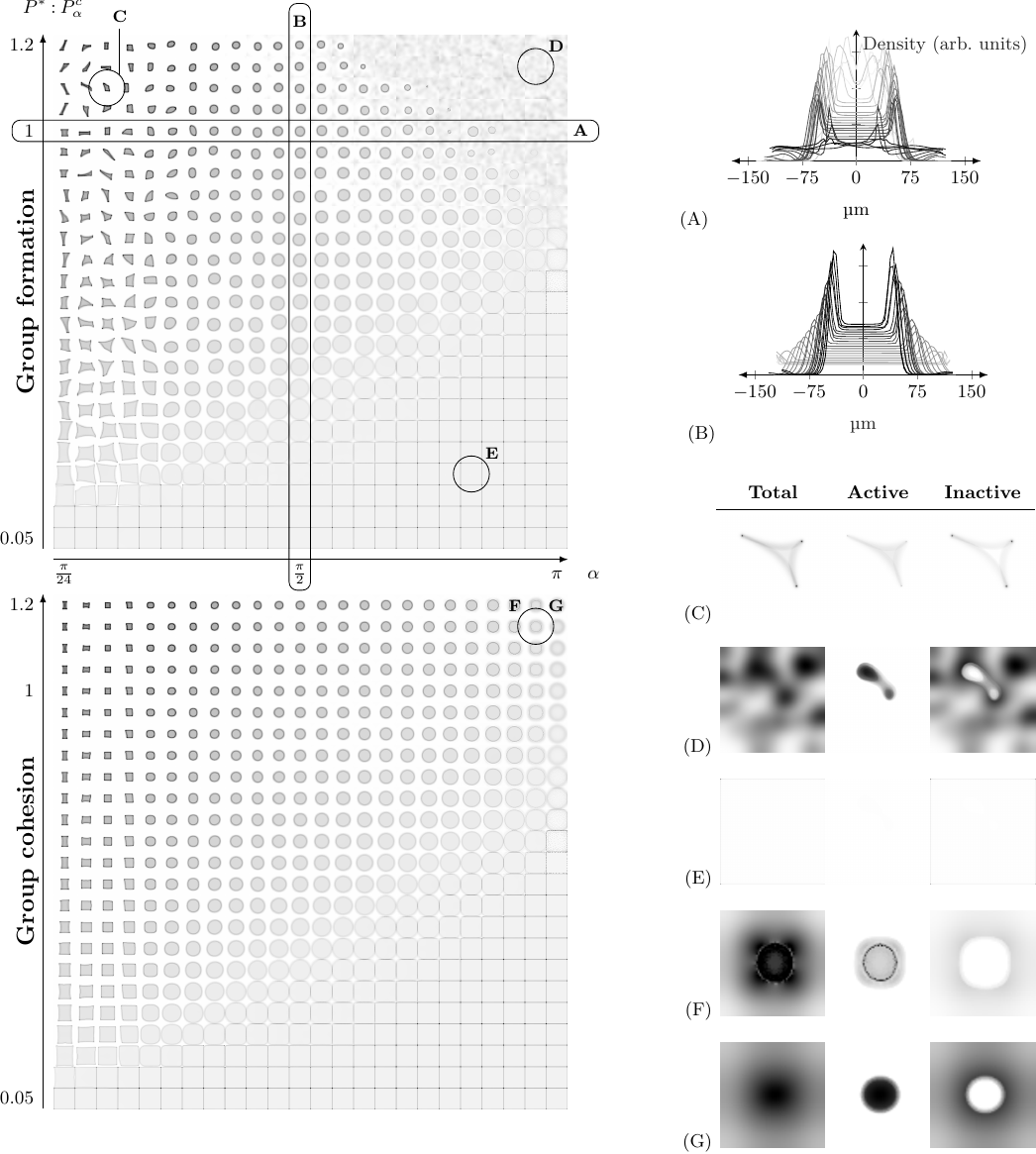}
    \caption{Left side: Phase diagram for group formation (top) and cohesion (bottom). PDE end-states for $\alpha$ ranging from $\frac{\pi}{24}$ to $\pi$ and $P^\ast\mathbin{:}P^c_{\alpha}$ ranging from \numrange{0.05}{1.2} are shown. Group \emph{formation} simulations start from random initialization, while \emph{cohesion} simulations start from annular end-state obtained using $\alpha = \frac{\pi}{2}$ and $P^\ast\mathbin{:}P^c_{\alpha} = 1$. Details: \textbf{(A)} examines the cross-sections along the row where $P^\ast\mathbin{:}P^c_{\alpha} = 1$. Darker lines for increasing $\alpha$. \textbf{(B)} examines the cross-sections along the column where $\alpha = \frac{\pi}{2}$. Darker lines for increasing $P^\ast$. \textbf{(C)} Detail from the low-angle high-threshold region: The group forms as a few cluster points and jets between them. \textbf{(D)} Detail from the high-angle high-threshold region: Particles never activate and groups do not form. \textbf{(E)} Detail from the low-threshold region: Particles are active everywhere but at domain boundaries. \textbf{(F)--(G)} long term high-angle high-threshold cohesion simulation  ($\alpha = \frac{23 \pi}{24}$ and $P^\ast\mathbin{:}P^c_\alpha = 1.15$, \SI{50000}{\second} and \SI{100000}{\second}, respectively). At a very long time scale, the central group collapses as particles move away from and lose contact with the cluster.}
    \label{fig:phasestuff}
\end{figure*}

This two-dimensional model parameter space was sampled as a $24\times24$ grid.  The vision parameter $\alpha$ ranges from $\frac{\pi}{24}$ to $\pi$ while $P^\ast\mathbin{:}P^c_\alpha$ varies from \numrange{0.5}{1.2}. Both agent-based and PDE simulations were run on the specified parameter values, keeping other parameters constant. (Only PDE results shown and discussed, here, as agent-based simulations performed equivalently and are redundant.)
Having previously established robustness with respect to spatial resolution, above, and in an effort to make computation time reasonable, the spatial resolution was lowered to $N_{\vec{x}}=64$. Similarly, the angular resolution was lowered to $N_\theta=30$, both without a noticeable impact on the observed pseudo-steady state. A visual summary of the results from various perspectives is shown in figure~\ref{fig:phasestuff}.

\subsubsection{Group formation}
The formation of an annular pattern for moderate parameter choices $\alpha=\frac\pi2$ and $P^\ast=P^c_\alpha$ was established, above. 
In figure~\ref{fig:phasestuff}\textbf{(A)}, we first examine the pattern formation for a fixed threshold and varying vision angle, $\alpha$. For narrow vision cones, we notice that the emerging pattern loses symmetry; the structure that forms essentially reduces to a small set of cluster points, defining the vertices of an often concave near-polytope. At the other end of the spectrum, at high $\alpha$-values, particles largely fail to activate (in part due to the increased $P^c_\alpha$ threshold) and are subject to regular diffusion, only. As a result, groups cease to form.

In contrast, figure~\ref{fig:phasestuff}\textbf{(B)} explores the dependency of the pattern on $P^\ast$ for $\alpha=\frac\pi2$, fixed. 
We observe that high perception values yield tighter groups while lower perception thresholds increase group radius, and eventually inflate the group against the domain boundaries. The cause, here, is that lower $P^\ast$ keeps particles activated and moving outwards with a lower amount of peers in front of them, allowing the particles to stretch the periphery.

Representatives from parameter space (highlighted in the top phase-diagram of figure~\ref{fig:phasestuff}) were then run on a higher spatial and angular resolution.

Figure~\ref{fig:phasestuff}\textbf{(C)} explores the low angle and high perception ratio; the separation between active and inactive particles suggests that predominantly, rather than an annular attractor, here the exchange of mass essentially happens between a few discrete point masses.

In figure~\ref{fig:phasestuff}\textbf{(D)}, we consider the region with both high angle and high perception ratio. Here, nearly all the particles are inactive resulting in diffusion-dominated dynamics: With the activation threshold being so high, agents have little chance of ever becoming active. This break down of group formation with high $\alpha$ and relatively high $P^\ast$ values is consistent with experimental results from \cite{lavergne2019group}; however, the exact phase transition seems to happen for slightly higher $\alpha$-values in the PDE model.

Figure~\ref{fig:phasestuff}\textbf{(E)} highlights the case for high alpha and low perception ratio. 
It suggests that agents are active anywhere in the interior of the domain. As a result, groups form along the boundaries, specifically the corners, where outward looking particles can become inactive.

\subsubsection{Group cohesion}
In contrast to pattern formation from random initialization, we now consider group cohesion. To study the cohesion, PDE simulations were run on the same parameter space, but this time all simulations were started from the $\alpha=\frac\pi2$ and $P^\ast=P^c_\alpha$ end-state annular configuration. The resulting phase-diagram can be found at the bottom left of figure \ref{fig:phasestuff}. 

Direct comparison of the group formation versus cohesion phase-portraits manifests two major differences. First,  observe the upper right-hand corner of both phase-portraits. While group formation does not occur for a large range of parameters, our cohesion results indicate that for some of the same parameters these patterns can persist. Familiar annular structures with sharp boundaries persist in many cases but the last few columns (near and complete surround vision); indeed, these distributions in the phase diagram are not stationary, yet, and continue to collapse and diffuse to uniformity on a much longer time scale, see figure~\ref{fig:phasestuff}\textbf{(F)--(G)}. 

Second, we notice that the top-left corner region of the phase diagram (low angle and high perception threshold) again manifests a breakdown of the annular pattern towards discrete vertices. However, in contrast to the group formation setting, here, the collapsing patterns retain a much larger degree of symmetry, typically collapsing to square or rectangular shapes aligned with the principal axes of spatial discretization. We believe this symmetry and alignment is primarily an artifact of the discretization and numerical schemes.

Finally, original results from \cite{lavergne2019group} suggest a lack of cohesion with low perception threshold and high alpha values (bottom right corner). Our results agree as the patterns inflate and are boundary-limited for the entire bottom-right parameter region; however, our phase-diagram does not exactly confirm the relatively narrow non-cohesive area postulated in \cite{lavergne2019group}.

\section{Discussion and Conclusions}
\label{sec:discussion}
In this paper we studied particle dynamics first proposed and analyzed in \cite{lavergne2019group}. Unlike classical swarming models, this model does not rely on pairwise attraction/repulsion and/or alignment, but only ``vision''-based sensing and selective forward motion on top of random drift. It was shown experimentally, in \cite{lavergne2019group}, that these dynamics, too, have the potential to lead to pattern formation and group cohesion. Here, we revisit an agent-based approach and propose a PDE-model for the numerical simulation of the particle dynamics. We introduce time-integrated density analysis for the discrete particle simulator and are thereby able to show that the two models are consistent. The models confirm an overall disk-shaped pattern emerging for moderate model parameters. This is consistent with the results reported in \cite{lavergne2019group}. Moreover, both the time-integrated agent-based densities and the PDE-based densities reveal an annular distribution within the disk, which was not previously reported in \cite{lavergne2019group}. This annular pattern was unnoticed in \cite{lavergne2019group} because it is much less apparent from discrete particle dynamics snapshots with low particle numbers ($n=75$ was used, there, for \emph{in vitro} and \emph{in silico} experiments). In contrast, our PDE model  considers the continuum limit. Our numerical experiments reveal that inactive particles mostly aggregate at the disk boundary, while active particles quickly traverse the interior of the group, confirming and explaining what was partially observed in \cite{lavergne2019group}. 

A possible limitation of our model (and the agent-based simulator in \cite{lavergne2019group}) is the issue of particle stacking.
Experimental particles had a diameter of \SI{4.28}{\micro\meter}. Meanwhile, our models ignore this size and therefore allow for (unrealistic) particles to overlap. This concern could be addressed by including a short-range repulsion term into both the agent-based and PDE-based models. However, we do not anticipate that this would substantially affect the nature of the results reported here.

\bibliography{references}

\end{document}